%% file: main.tex
\documentclass[12pt]{article}

\usepackage{textcomp}
\usepackage{multirow}
\usepackage{amsmath} 
\usepackage{graphicx} 
\usepackage{setspace} 
\usepackage{authblk}
\usepackage{tabularx} 
\usepackage{booktabs} 
\usepackage{cite}
\usepackage[a4paper,
			bindingoffset=0.2in,
			left=1in,
			right=1in,
			top=1in,
			bottom=1in,
			footskip=0.25in]{geometry}

\title{When Trace Water Dominates: Hydration–Mediated Dielectric and Transport Behaviour in BiFeO$_3$}

\author[1,2]{Subir Majumder}
\author[2]{Gilad Orr}
\author[1,*] {Paul Ben Ishai}
\affil[1]{THz and Dielectric Science Lab., Dept. of Physics, Ariel University, Ariel, Israel}
\affil[2]{Crystal Physics Lab., Dept. of Physics, Ariel University, Ariel, Israel}
\affil[*]{Corresponding Author: paulbi@ariel.ac.il} 
\begin{document}

\maketitle
	

\begin{abstract}
Traces of water can profoundly alter the dielectric response of functional oxides, yet such effects have remained largely unrecognized in systems where colossal dielectric behaviour has been widely reported. Here, we investigate the impact of sub-percent hydration ($<$1 wt\%) on the dielectric relaxation, charge transport, and interfacial polarization properties of porous BiFeO$_3$ ceramics. Broadband dielectric spectroscopy reveals, in the hydrated state, a dominant relaxation process characterized by an anomalously large dielectric strength ($\Delta\varepsilon \approx$ 10$^4$-10$^5$) and a pronounced saddle-point deviation from Arrhenius dynamics, indicative of non-Arrhenius relaxation behaviour in a porous oxide system. These features appear only in the hydrated state and vanish upon dehydration, while the intrinsic activation barriers governing the thermally activated relaxation timescale remain comparable. Comparison with hydration-controlled dielectric responses in layered clay minerals shows that similar qualitative deviations can emerge in BiFeO$_3$ with nearly fifteen-fold lower water content, underscoring the effectiveness of confined water at grain boundaries, pore surfaces, and internal interfaces. Together, these results demonstrate that trace, confined water can make a major extrinsic contribution to dielectric and transport anomalies in porous oxide ceramics. The use of dehydration-controlled dielectric cycling provides a practical diagnostic framework for reassessing colossal dielectric responses, Maxwell-Wagner-type effects, and hydration-induced phenomena in functional oxide materials.
\end{abstract}

\vspace{10pt}

\maketitle

\section{Introduction}
Dielectric relaxation phenomena in functional oxides\cite{Zhao2013,Singh2023, Li2018} are widely used to probe charge dynamics, defect states, and interfacial polarization mechanisms. In many technologically relevant oxides, particularly porous and polycrystalline ceramics \cite{Zhao2013,Singh2023}, broadband dielectric spectroscopy reveals complex temperature and frequency dependent responses that are often interpreted in terms of intrinsic dipolar relaxations, defect hopping, or Maxwell-Wagner-type interfacial effects \cite{Anjeline2021,Wang2017}. However, disentangling intrinsic mechanisms from extrinsic contributions remains a persistent challenge, especially when weakly bound species are present at interfaces, grain boundaries, or pore surfaces. Water is often one such factor. While the influence of hydration is well documented in soft materials\cite{Vasilyeva2014,Polansky2017,Mitsari2017,Bidadi1988}, including polymers and clay minerals, its role in dense oxide ceramics is often assumed to be negligible when the water content is small. This assumption is reinforced by thermogravimetric analyses that typically report mass losses well below a few weight percent, leading to the perception that sub-percent hydration cannot significantly contribute to macroscopic dielectric or transport behaviour. Recent studies on hydrated layered silicates and clay minerals have challenged this perspective by demonstrating that water can induce collective dielectric relaxations characterized by anomalously large dielectric strengths and non-Arrhenius, saddle-point-like relaxation dynamics\cite{Mitsari2017, Sjostrom2008,Sasaki2020,Cerveny2011}. In these systems, water confined within interlayer galleries\cite{Vasilyeva2014,Mitsari2017,Jeevanandam1998} gives rise to temperature-dependent connectivity and cooperative polarization, producing dielectric responses that cannot be explained by independent dipoles or simple activated processes. Whether similar physics can emerge in dense oxide ceramics, where water is confined not in extended layers but in pores and grain boundaries, remains an open and largely unexplored question. BiFeO$_3$ is a well-studied functional oxide with robust structural stability\cite{Lin2014}, it exhibits multiple dielectric relaxation processes across wide temperature and frequency ranges\cite{Lin2014,Markiewicz2011,Chybczynska2016,Zheng2020}. At the same time, porous BiFeO$_3$ ceramics\cite{Chybczynska2016,Narayan2024} can host small amount of confined water at internal surfaces, making them an ideal testbed to examine whether trace hydration can induce collective dielectric and transport phenomena.
To address this question, we combine broadband dielectric spectroscopy, relaxation-time analysis, thermogravimetric quantification, and dc conductivity measurements across controlled heating cycles in porous BiFeO$_3$ ceramics. By systematically comparing hydrated and dehydrated states, this approach enables a direct assessment of the role of confined water in shaping dielectric and transport responses.

\section{Materials and Methods}
Polycrystalline BiFeO$_3$ ceramics were synthesized by a conventional solid-state reaction route\cite{Majumder2025,Orr2022,Orr2020Sintering} using stoichiometric precursor oxides, followed by calcination at 400$^o$C for 2 hours and sintering at  880$^o$C for 08 minutes to obtain phase-pure\cite{Wang2004}, porous ceramics. These samples were pressed at a uni-axial pressure of 100MPa into pellets of 12 mm in diameter. XRD was carried out on a Rigaku SmartLab SE diffractometer using Cu-K$\alpha$ radiation ($\lambda$ = 1.5406 $\mathring{A}$) with $\Delta\theta$ = 0.02$^o$. Rietveld refinement was performed in Profex\cite{Doebelin2015} software for phase identification, lattice parameter calculation, and quantitative phase analysis\cite{Orr2021}, using COD\cite{Grazulis2009,Grazulis2012} database entries. Microstructure was examined with a TESCAN Maia3 FE-SEM at 7 kV. The as-sintered samples were stored under ambient laboratory conditions prior to measurement, such that the initial heating cycle probes the naturally hydrated state. Dehydration was achieved in situ by holding the sample at 250$^o$C for 15 min, after which subsequent measurements reflect the dehydrated response [see the common inset of Fig. 3]. Broadband dielectric spectroscopy measurements were performed over the temperature range -130$^o$C to 250$^o$C and frequency range 0.1 Hz to 10 MHz using a Novocontrol BDS 80 dielectric spectrometer with a Quattro temperature control system based on liquid Nitrogen. Silver electrodes were deposited on opposite faces of the pellet to ensure an Ohmic contact and a defined geometric capacitance of the sample\cite{Novocontrol2000}. Dielectric spectra were recorded during controlled heating protocols with defined stabilization times at each temperature to ensure thermal equilibrium. The same sample and electrode configuration were used throughout all heating cycles to ensure direct comparability between hydrated and dehydrated states.

\section{RESULTS AND DISCUSSION}

\subsection{Structural Characterization and Hydration State}

X-ray diffraction confirms single-phase rhombohedral BiFeO$_3$ [Fig. 1(a)]. Rietveld refinement did not show any secondary phases within the detection limit of the measurement. Scanning electron micrograph reveals a porous microstructure with uniform compositional contrast [Fig. 1(b)]. The presence of open pores provides internal surfaces and interfaces that may host weakly bound species under ambient conditions and lead to the absorption of atmospheric moisture.
To quantify the amount of such species, present in the ceramic, thermogravimetric analysis was performed. The TGA curve (Fig. 2) recorded a mass loss of $~$0.9 wt\% between 40$^o$C and 90$^o$C. This indicates weakly bound or confined water associated with internal surfaces, pore walls, and maybe grain boundaries of the porous ceramic. Although most of the mass loss occurs below 90$^o$C, the dielectric measurements extend to higher temperatures where the dehydration process continues to evolve and influence the interfacial dielectric response. Although the total water content is small, confined water at pore surfaces and grain boundaries can influence dielectric relaxation\cite{Vasilyeva2014,Gutina2003}. 

\begin{figure}[htbp]
    \centering
    \includegraphics[width=0.98\textwidth]{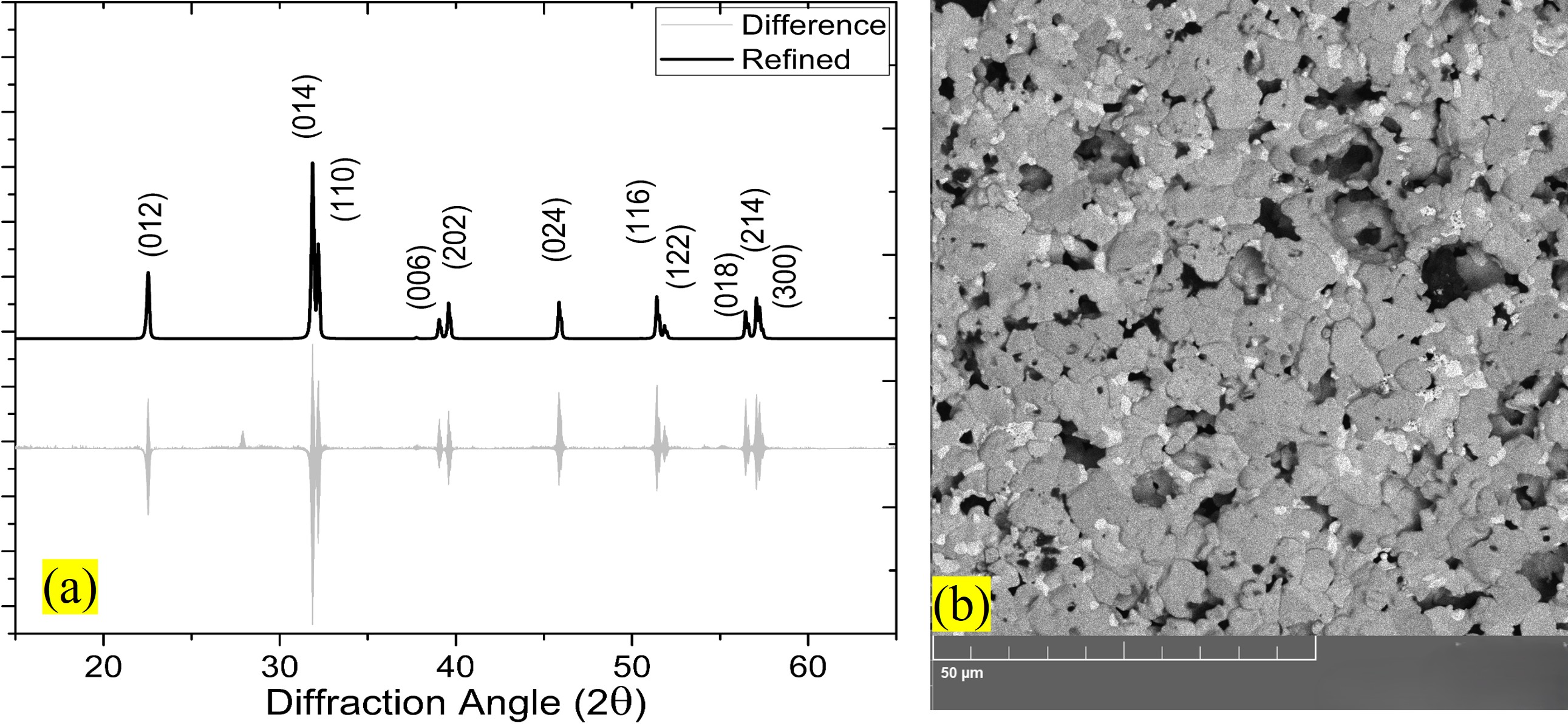}
    \caption{X-ray diffraction pattern and microstructure of the BiFeO$_3$ ceramic. (a) XRD pattern confirming single-phase rhombohedral structure. (b) SEM micrograph showing the porous microstructure of the sintered ceramic.}
    \label{fig:xrd-sem}
\end{figure}

\begin{figure}[htbp]
    \centering
    \includegraphics[width=0.55\textwidth]{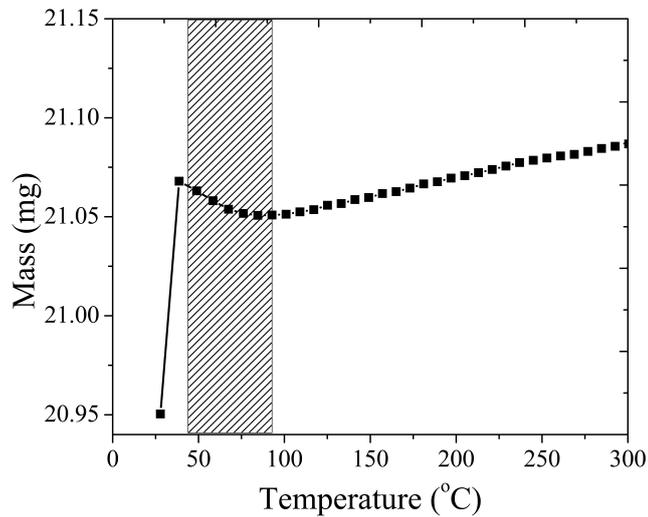}
    \caption{Thermogravimetric analysis (TGA) of the BiFeO$_3$ ceramic showing a mass loss of $\sim$0.9 wt\% between 40$^\circ$C and 90$^\circ$C, attributed to weakly bound or confined water associated with internal surfaces, pore walls, and grain boundaries.}
    \label{fig:TGA}
\end{figure}

\subsection{Dielectric Loss Landscape and Identification of Relaxation Processes}
Structural and microstructural stability across heating cycles ensures that changes in dielectric and transport properties originate from extrinsic effects rather than irreversible structural modification. Broadband dielectric spectra recorded during the two heating cycles reveal three relaxation processes ($P1-P3$) in each cycle together with a low-frequency dc conductivity contribution (Fig. 3a,b). 
For clarity, the same process labels ($P1-P3$) are retained for both heating cycles, although their temperature ranges and spectral characteristics differ between the hydrated and dehydrated states. Process $P1$ appears predominantly at low temperatures and exhibits modest dielectric strength typical of localized intrinsic relaxations. In the hydrated cycle it spans a broad temperature range within the measurement window, while after dehydration it becomes restricted to the lower-temperature region. Process $P2$ occurs over an intermediate to high temperature range and constitutes the dominant relaxation in the spectra. In the hydrated state it displays a characteristic saddle-like temperature dependence of the relaxation time, whereas after dehydration it reverts to conventional Arrhenius behaviour. Process $P3$ appears at the low temperatures in the hydrated cycle and shifts toward higher temperatures after dehydration, where it increasingly overlaps with the low-frequency dc conductivity contribution.

The spectra were analysed using a superposition of Havrilak-Negami functions\cite{Kremer2002} and dc conductivity term: 

\begin{equation}\label{eq:Havriliak-Negami}
\varepsilon_m^*(\omega) = \varepsilon_\infty + \sum_{n=1}^{3} \frac{\Delta \varepsilon_n}{\Bigl[1 + (i \omega \tau_n)^{\alpha_n}\Bigr]^{\beta_n}} 
+ \frac{\sigma_{\rm dc}}{i \omega \varepsilon_0}
\end{equation}

where $\Delta\varepsilon_n$ is the dielectric strength of process $n$, $\tau_n$ is its characteristic relaxation time, 0$< \alpha_n$, $\beta_n\leq$1 are the shape parameters, $\varepsilon_0$ is the permittivity of free space and $\varepsilon_{\infty}$ is the high frequency limit of the permittivity. The spectra were modelled and analysed using a bespoke model fitting software, DATAMA\cite{Axelrod2004}, implemented in MATLAB$\copyright$. 
While the number of relaxation processes remains unchanged in the two heating cycles, dielectric strength changes strongly, whereas the activation energies remain similar. This behaviour indicates a strong hydration dependence of the dielectric response, with process $P2$ being particularly sensitive to the presence of confined water.

\begin{figure}[htbp]
    \centering
    \includegraphics[width=0.95\textwidth]{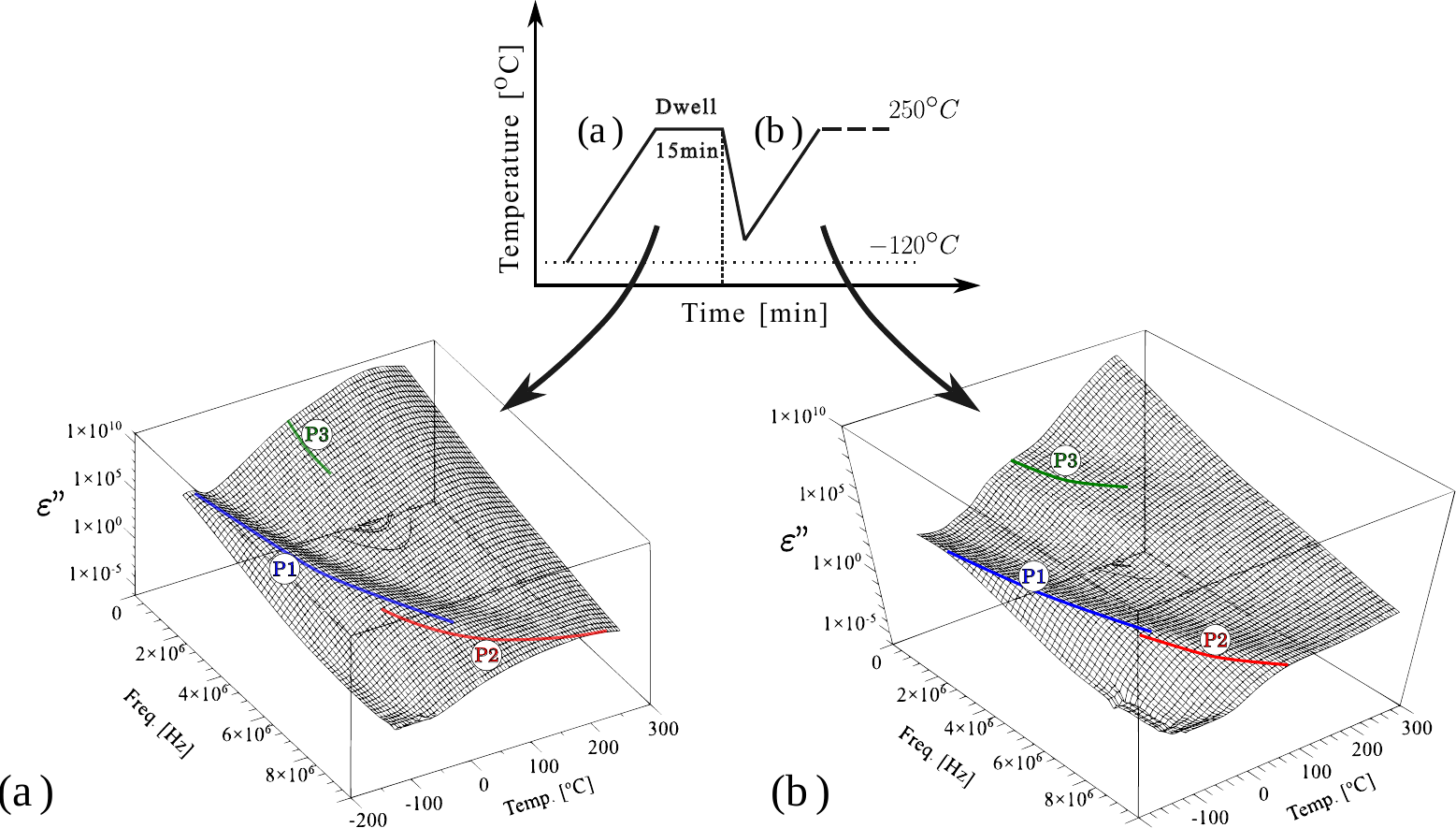}   
    \caption{Three-dimensional dielectric loss landscapes $\varepsilon''(T,f)$ measured during (a) the first heating cycle (hydrated state) and (b) the second heating cycle after dehydration. Three relaxation processes ($P1-P3$) are observed in both cycles and their stretch are marked in colour. Inset: schematic illustration of the heating-cycle protocol used to compare hydrated and dehydrated states.}
    \label{fig:dielectric-landscape}
\end{figure}

\subsection{Relaxation Dynamics: Arrhenius and Saddle-Point Behaviour}

The temperature dependence of the characteristic relaxation times, extracted from the spectral fits, is shown in Fig. 4. Processes $P1$ and $P3$ follow approximately Arrhenius behaviour over the accessible temperature range in both heating cycles, with activation energies in the range 33-40 kJ/mol (Table 1). The similarity of these activation energies before and after dehydration indicates that the underlying relaxation dynamics remain largely unchanged and are associated with defect-related charge hopping in the lattice (as discussed later, e.g., Fe$^{2+}$/Fe$^{3+}$ small-polaron hopping) or interfacial polarization processes typical of oxide ceramics.

Process $P2$ exhibits distinctly different behaviour. In the hydrated cycle the relaxation time deviates strongly from Arrhenius behaviour. It initially decreases with increasing temperature, reaches a minimum, and subsequently increases again at higher temperatures, producing a characteristic non-monotonic (saddle-like) dependence. After dehydration, this anomalous behaviour disappears and $P2$ reverts to conventional Arrhenius behaviour with an activation energy of approximately 88 kJ/mol. The temperature and frequency ranges, activation energies, and physical assignments of all processes identified in the two heating cycles are summarized in Table 1. The disappearance of the saddle-type relaxation after dehydration indicates that the anomalous dynamics observed in the hydrated state arise from an additional mechanism associated with confined water rather than from intrinsic lattice relaxation.

\begin{figure}[htbp]
    \centering
    \includegraphics[width=0.95\textwidth]{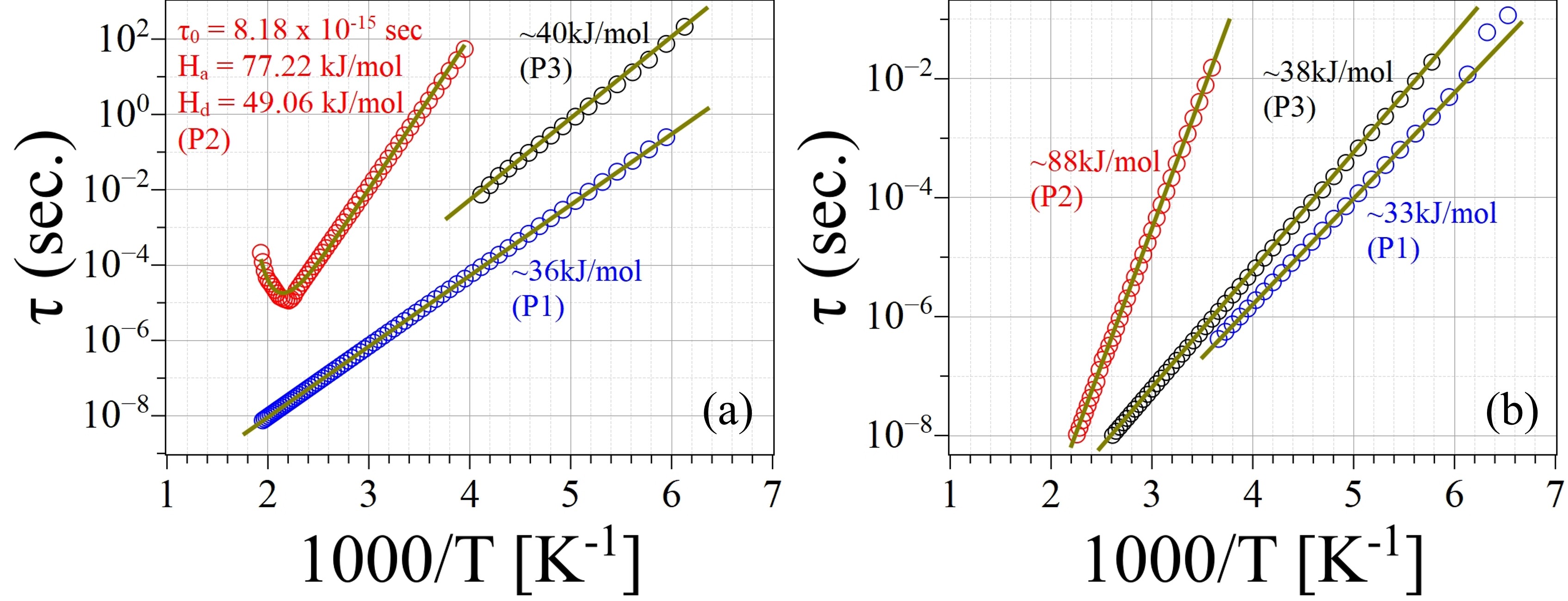}  
    \caption{Arrhenius plots of the relaxation times for processes $P1$, $P2$, and $P3$ shown for (a) the first heating cycle (hydrated state) and (b) the second heating cycle (dehydrated state). Activation energies obtained from linear regions are indicated. A saddle-like deviation from Arrhenius behaviour is observed only in the hydrated state.}
    \label{fig:Arrhenius}
\end{figure}

\begin{table}[htbp]
\centering
\caption{Summary of dielectric relaxation processes observed in the two thermal cycles.}

\footnotesize
\setlength{\tabcolsep}{3pt}
\renewcommand{\arraystretch}{1.15}

\begin{tabular}{|c|c|c|c|c|c|p{4.1cm}|}
\hline

Cycle & Proc. & Temp. ($^o$C) 
& \begin{tabular}{c} Frequency \\ (Hz) \end{tabular}
& \begin{tabular}{c} $E_a$ \\ (kJ/mol) \end{tabular}
& \begin{tabular}{c} $E_a$ \\ (eV) \end{tabular}
& Physical Assignment \\
\hline

\multirow{3}{*}{\begin{tabular}{c}1st \\(hydrated) \end{tabular}}
& P1 
& \begin{tabular}{c} (-)105--\\(+ )240 \end{tabular}
& \begin{tabular}{c} 0.64--\\2.1 $\times$ 10$^{7}$ \end{tabular}
& 36 
& 0.37 
& Localized polaron hopping or dipolar relaxation associated with lattice defects (e.g., Fe$^{2+}$/Fe$^{3+}$ hopping) \\
\cline{2-7}

& P2 
& \begin{tabular}{c}(-)20--\\(+ )245 \end{tabular}
& \begin{tabular}{c} 2.9 $\times$ 10$^{-3}$--\\1.4 $\times$ 10$^{4}$ \end{tabular}
& \begin{tabular}{c}$H_a=77.2$, \\$H_d=49.1$ \end{tabular}
& \begin{tabular}{c} $H_a=0.80$, \\$H_d=0.51$ \end{tabular}
& Defect-mediated relaxation involving thermally activated defect formation (water/oxygen-vacancy related saddle-point dynamics) \\
\cline{2-7}

& P3 
& \begin{tabular}{c}(-)110--\\(-)30 \end{tabular}
& \begin{tabular}{c} 7.5 $\times$ 10$^{-4}$--\\21 \end{tabular}
& 40 
& 0.41 
& Interfacial/space-charge polarization associated with grain boundaries or electrode interfaces \\
\hline

\multirow{3}{*}{\begin{tabular}{c}2nd \\(dehydrated) \end{tabular}}
& P1 
& \begin{tabular}{c}(-)120--\\(+ )0 \end{tabular}
& \begin{tabular}{c} 1.4--\\3.8 $\times$ 10$^{5}$ \end{tabular}
& 33 
& 0.34 
& Residual localized dipolar hopping after dehydration \\
\cline{2-7}

& P2 
& \begin{tabular}{c} (+)5--\\(+ )170 \end{tabular}
& \begin{tabular}{c} 0.1--\\1.5 $\times$ 10$^{7}$ \end{tabular}
& 88 
& 0.91 
& Thermally activated defect relaxation dominated by intrinsic lattice defects \\
\cline{2-7}

& P3 
& \begin{tabular}{c} (-)100--\\(+ )110 \end{tabular}{c}
& \begin{tabular}{c}8.3--\\1.6 $\times$ 10$^{7}$ \end{tabular}{c}
& 38 
& 0.39 
& Weak interfacial polarization due to remaining charge trapping centres \\
\hline

\end{tabular}
\end{table}

The anomalous relaxation behaviour of process $P2$ in the hydrated state can be described using the saddle-point relaxation model proposed for confined systems. Within this framework the temperature dependence of the relaxation time is expressed as\cite{Vasilyeva2014,Sjostrom2008}

\begin{equation}\label{eq:saddle}
\tau = \tau_0 \exp\left[ \frac{H_a}{kT} + C \exp\left( -\frac{H_d}{kT} \right) \right]
\end{equation}

where $H_a$ represents an effective activation barrier governing dipolar relaxation with a single activation energy, $H_d$ represents the energy of local defect-formation need to permit the dipole relaxation. As the probability of a defect forming in the immediate vicinity of the dipole is governed by Boltzmann statistics, it gives rise to a temperature dependent exponential term. $C$ is a dimensionless collective parameter inversely proportional to the maximum concentration of such defects  within the confined system. Curve fitting yields $H_a$ = 77.2 kJ/mol  and $H_d$ = 49.1 kJ/mol, with a minima at $T_s\approx$465 $K$ ($\approx$ 192$^o$C). The similarity between the intrinsic activation barrier $H_a$ obtained from the saddle-point fit and the Arrhenius activation energy observed for $P2$ after dehydration (88 kJ/mol) indicates that hydration does not significantly modify the fundamental hopping barrier. Instead, hydration introduces an additional population of defects, whose presence enhances the dielectric strength and produces the observed non-Arrhenius relaxation behaviour. Within the Ryabov formulation\cite{Ryabov2004,BenIshai2006}, such a minimum exists if and only if the collective parameters satisfy the necessary and sufficient condition $C > \frac{H_a}{H_d}$ (derivation in Supplementary Material). Given the inverse proportionality of $C$ to the defect concentration and that the enthalpy, $H_d$, is reminiscent of the activation energy for interfacial water, this condition suggests that one water molecule in the vicinity of 2 deep electronic trap states is sufficient to induce saddle-like behaviour. In this framework the emergence of the relaxation-time minimum reflects the competition between thermally activated relaxation and the temperature-dependent population of necessary defects within the confined hydrated environment, leading to a saddle behaviour in $\tau(T)$. In the dehydrated state this collective contribution disappears (C$\rightarrow$0), and the expression reduces to the conventional Arrhenius form governed by the intrinsic barrier.  The coincidence of $T_s$ with the saturation and subsequent reduction of $\Delta\varepsilon$ is consistent with a common origin of the enhanced dielectric strength and the non-Arrhenius relaxation dynamics. Thus, hydration does not measurably alter the intrinsic activation barrier but is correlated with the emergence of a regime in which anomalously large polarization and saddle-point relaxation appear concurrently. The influence of hydration on charge transport is further reflected in the temperature dependence of the dc conductivity discussed below.

\subsection{Hydration-Dependent Dielectric Strength}
Further insight into the nature of the relaxation processes can be obtained from the temperature dependence of the dielectric strength $\Delta\varepsilon$ extracted from the spectral fits. Figure 5 shows the evolution of $\Delta\varepsilon$ associated with processes $P1$, $P2$, and $P3$ during the two heating cycles.

\begin{figure}[htbp]
    \centering
    \includegraphics[width=0.95\textwidth]{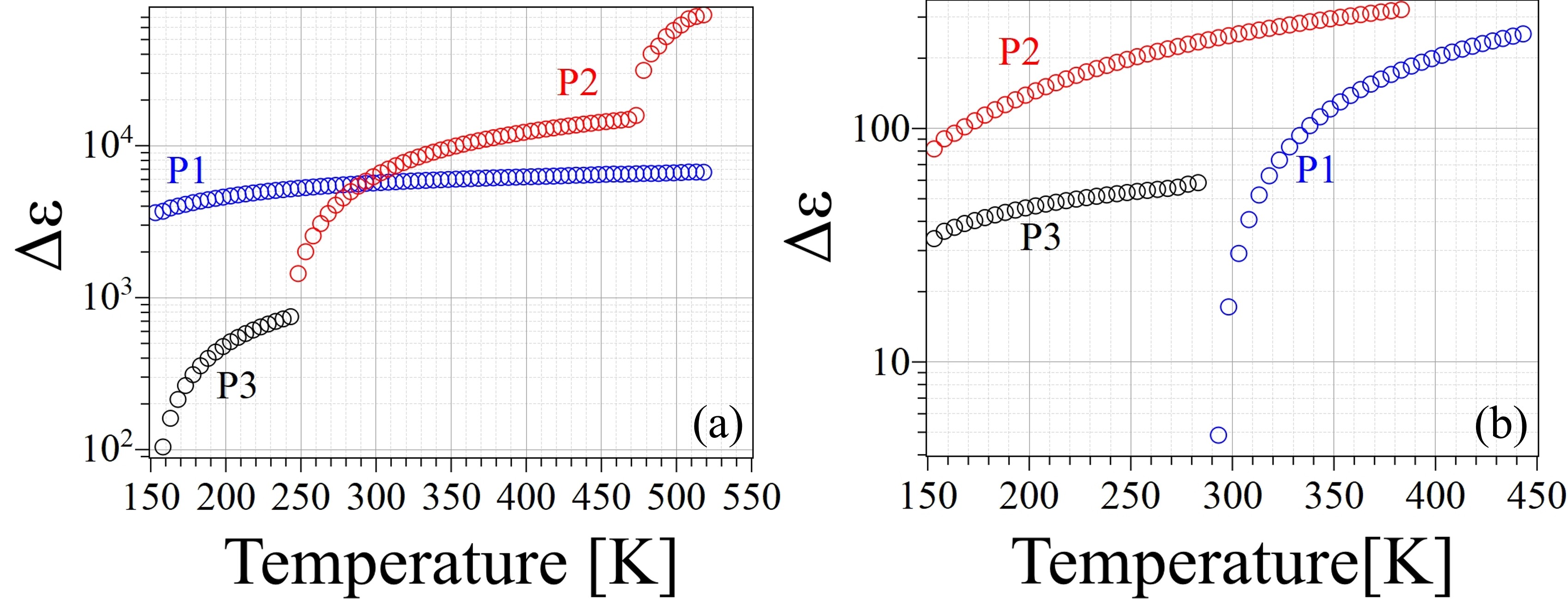}   
    \caption{Temperature dependence of the dielectric strength $\Delta\varepsilon$ associated with relaxation processes $P1$, $P2$, and $P3$ for the first (hydrated) and second (dehydrated) heating cycles.}
    \label{fig:dielectric-strength}
\end{figure}

In the hydrated state, the dominant process $P2$ exhibits an exceptionally large dielectric strength, increasing from approximately $1.4\times 10^3$ to $7.3\times 10^4$ between -20$^o$C and 245$^o$C. Such values are unusually high for intrinsic dipolar or defect-related mechanisms in dense oxide ceramics and are difficult to reconcile with purely intrinsic contributions. The remaining processes display comparatively modest dielectric strengths consistent with localized intrinsic relaxations. The magnitude of the dielectric strength can be expressed phenomenologically as

\begin{equation}\label{eq:dielectric-strength}
\Delta \varepsilon = \frac{N_{\mathrm{eff}} \mu_{\mathrm{eff}}^{2}}{kT} \, g
\end{equation}

where $N_{eff}$  is the effective density of active dipoles, $\mu_{eff}$  their effective dipole moment, and $g$ is the Kirkwood correlation factor\cite{Kremer2002,Bottcher1973,Frohlich1949} that accounts for the enhancement of the dipole moment by interaction with neighbouring dipole and the dielectric background. Within this framework, hydration is associated with the enhanced participation of dipolar contributions that increase the effective dipole density and enhance orientational correlations. Such contributions can produce a polarization response far exceeding that expected from intrinsic defect dipoles alone. Furthermore, one notes that the dielectric strength increases with temperature for all processes. This suggests that the effective dipole concentration, $N_{eff}$, coupled with the corelation factor, $g$, must increase faster that the temperature depreciation implicit in equation (3).
After dehydration, the dielectric strengths of all the processes were strongly reduced. This behaviour indicates that the large polarization response observed in the hydrated state is closely linked to the presence of confined water at internal surfaces and grain boundaries.
The temperature at which the relaxation time of $P2$ reaches its minimum ($T_s \approx$465$K$ ($\approx$192$^o$C)) coincides with the temperature region where $\Delta\varepsilon$ stabilizes and subsequently decreases. This correspondence suggests that the enhanced polarization and the non-Arrhenius relaxation dynamics originate from the same hydration-dependent mechanism. Hydration therefore introduces an additional polarization contribution without significantly modifying the intrinsic activation barrier of the relaxation process.

\subsection{DC Conductivity and Hydration-Controlled Transport Enhancement}
The temperature dependence of the dc conductivity $\sigma_{dc}$ measured during the two heating cycles is shown in Fig. 6. In the hydrated state the conductivity exhibits a non-Arrhenius, sigmoidal behaviour, increasing by several orders of magnitude with temperature up to a maximum near $\sim$473$K$($\approx$200 $^o$C), followed by a decrease at higher temperature as hydration-mediated conduction pathways become progressively destabilized\cite{Kreuer2003}. This non-monotonic response indicates the presence of an additional hydration-dependent contribution to charge transport that is active only over a limited temperature range.

\begin{figure}[htbp]
    \centering
    \includegraphics[width=0.55\textwidth]{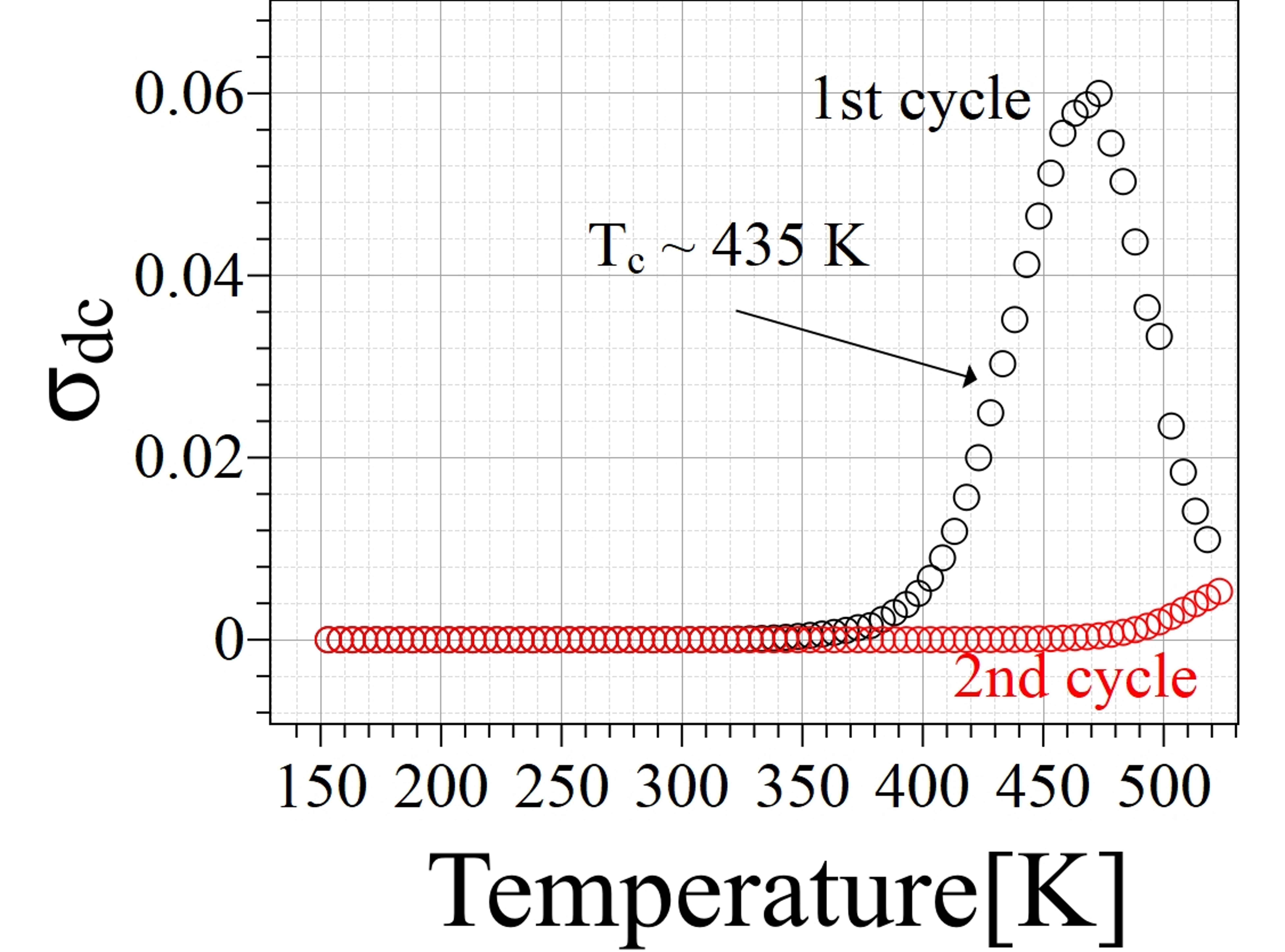}   
    \caption{Temperature dependence of the dc conductivity $\sigma_{\rm dc}$ measured during the first (hydrated, black symbols) and second (dehydrated, red symbols) heating cycles. The hydrated state exhibits a sigmoidal temperature dependence with a characteristic threshold temperature followed by a high-temperature decrease, whereas the dehydrated state shows a smooth monotonic increase over the entire temperature range.}
    \label{fig:dc-conductivity}
\end{figure}

The rapid increase of conductivity during the first heating cycle above approximately $\sim 368$ $K$ ($\approx 95^\circ$C) suggests the progressive development of long-range connectivity between hydration-assisted conductive regions. Confined water molecules can form transient hydrogen-bond networks that facilitate charge transport through proton transfer or polaron-assisted hopping along hydration channels. At lower temperatures, these conductive regions remain spatially isolated, whereas increasing temperature promotes their interconnection and the formation of continuous transport pathways.

The conductivity behaviour in the hydrated state was analyzed within a percolation framework. The dc conductivity exhibits percolation-type behaviour with a threshold temperature\cite{Kirkpatrick1973,Stauffer2018,Broadbent1957} $T_c \approx 435$ $K$ ($\approx 162^\circ$C), separating regimes below and above the connectivity transition. In this formalism the conductivity follows\cite{Kirkpatrick1973, Efros1976,orr2018high,Straley1977,Webman1975}
\[
\sigma = \sigma_M (T - T_c)^t, \quad T > T_c
\]
\[
\sigma = \sigma_D (T_c - T)^{-q}, \quad T < T_c
\]

where $\sigma_M$ represents the effective conductivity when all conductive bonds are present and $\sigma_D$ corresponds to the effective conductivity when only conductive bonds contribute. The two regimes are related through
\[
\sigma(T_c) = \sigma_M \left( \frac{\sigma_D}{\sigma_M} \right)^s
\]
with the critical exponents satisfying the scaling relation
\[
q = t \left( \frac{1}{s} - 1 \right).
\]

The log-log representation used to determine the critical exponents is shown in Fig.7(a), where the conductivity data for both $T > T_c$ and $T < T_c$ are analyzed. The analysis yields $t \approx 0.26$ and $q \approx 0.22$. Using the scaling relation above, the corresponding exponent is $s \approx 0.55$. The exponent $q$ characterizes the conductivity behaviour below the threshold temperature\cite{Efros1976} and reflects charge transport through disconnected hydration-assisted clusters. The relatively small value of $q$ indicates that conductivity below $T_c$ is dominated by localized hopping between weakly connected conductive regions prior to the establishment of long-range connectivity. The relatively small value of $t$ indicates a gradual development of connectivity above the threshold, reflecting a thermally driven rather than purely geometric percolation process associated with hydration-assisted pathways. The exponent $s$ describes the relative contribution of the conducting and weakly conducting regions at the transition\cite{Efros1976,Bergman1977} and reflects the strong contrast between hydration-assisted conductive regions and the surrounding matrix. These results support the interpretation that the enhanced conductivity in the hydrated state arises from the formation of a heterogeneous network of hydration-mediated conduction pathways.

At higher temperatures, the conductivity decreases despite increasing thermal energy, indicating progressive disruption of this network as dehydration proceeds. The sigmoidal conductivity profile therefore reflects the formation and subsequent collapse of hydration-assisted transport pathways during heating.

In contrast, the dehydrated state shows a smooth monotonic increase characteristic of intrinsic thermally activated transport in oxide ceramics. The absence of the sigmoidal feature after dehydration indicates that the additional conductivity contribution observed during the first heating cycle originates from confined water. The Arrhenius representation of the second heating cycle is shown in Fig.7(b), where the conductivity is plotted as a function of $1000/T$. For temperatures above 363 K ($\approx 90^\circ$C) the data follow a linear dependence, yielding an activation energy of approximately $E_a \approx 110.18$ kJ/mol ($\approx 1.14$ eV), consistent with thermally activated hopping through localized defect states in the dehydrated lattice.

The conductivity crossover occurs in the same temperature range where the dielectric strength $\Delta \varepsilon$ reaches its maximum and the saddle-point relaxation associated with process $P2$ develops, indicating a direct link between the anomalous dielectric response and the hydration-assisted conduction pathways.

Taken together, the dielectric and transport results show that trace confined water does not significantly modify the intrinsic activation energies of the host lattice but introduces an additional hydration-dependent contribution to charge transport. This contribution becomes prominent within a limited temperature interval where hydration-mediated connectivity develops and is progressively destabilized at higher temperature. Once dehydration occurs, the system reverts to intrinsic thermally activated behaviour characterized by Arrhenius-type conductivity and weak dielectric response.

\begin{figure}[htbp]
    \centering
    \includegraphics[width=0.65\textwidth]{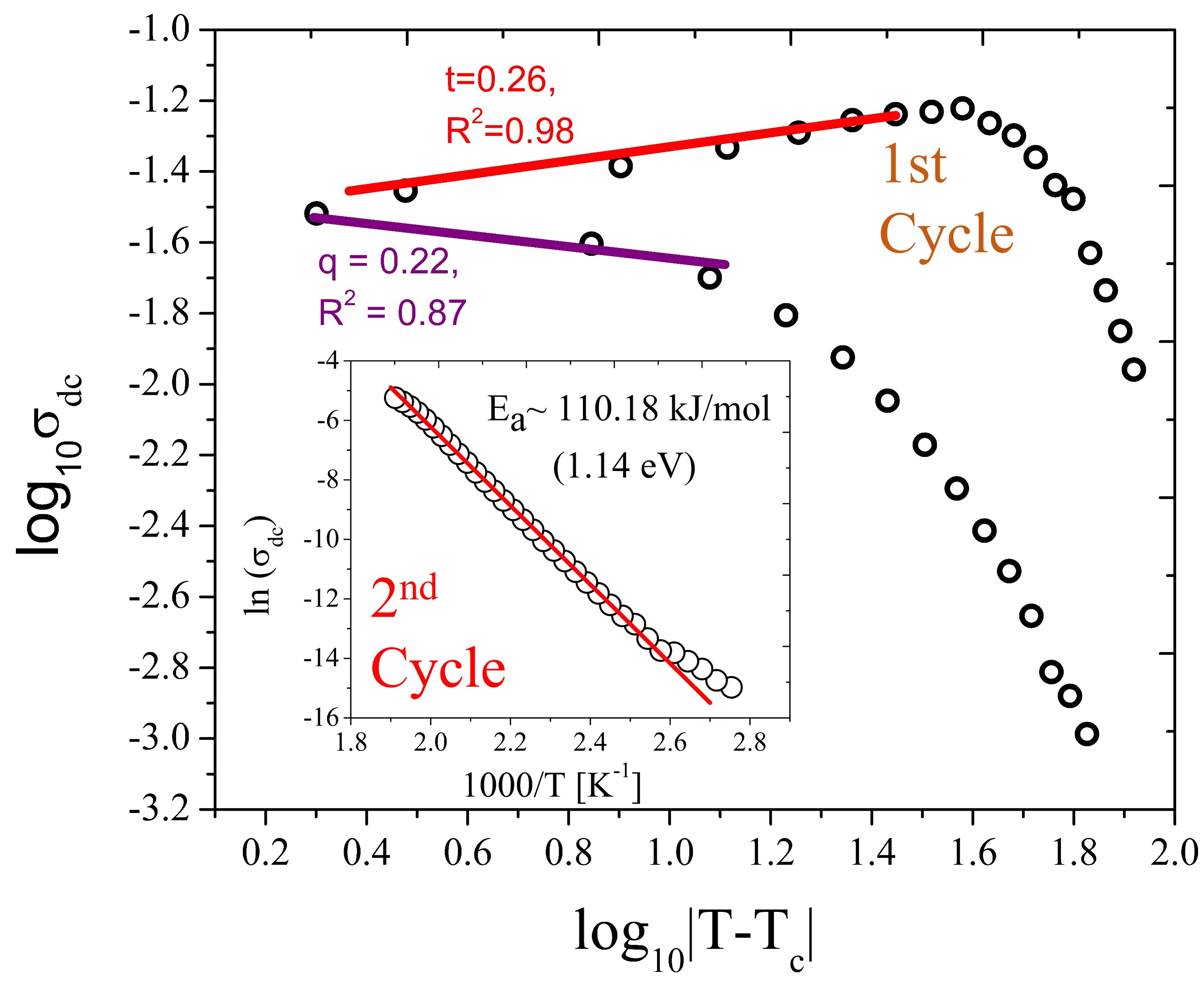}   
    \caption{Percolation and Arrhenius analysis of dc conductivity. The main panel shows the log--log plot of $\sigma_{\rm dc}$ versus $T - T_c$ for the first heating cycle ($T < T_c$ and $T > T_c$), used to obtain the critical exponents $s$ and $t$. The inset shows the Arrhenius plot of $\sigma_{\rm dc}$ versus $1/T$ for the second heating cycle, yielding $E_a = 0.91$~eV.}
    \label{fig:percolation-arrhenius}
\end{figure}

\subsection{Intrinsic Hopping and Hydration-Mediated Polarization}
The experimental observations described above establish a consistent sequence: the hydrated state exhibits a large dielectric strength, saddle-point relaxation dynamics, and enhanced conductivity, whereas after dehydration the relaxation processes revert to simple Arrhenius behaviour with strongly reduced dielectric strength. Thermally activated relaxations with similar activation energies ($\sim$0.25–0.4 eV) are widely reported in Fe-based oxides and are commonly attributed to Fe$^{+2}$/Fe$^{+3}$ electronic hopping\cite{Lin2014,Markiewicz2011,Ikeda2005}. The intrinsic hopping processes responsible for relaxations such as $P1$ provide the electronic timescale of the system, while hydration primarily modifies the interfacial polarization that gives rise to the large dielectric response. In BiFeO$_3$ ceramics such activation energies have previously been associated with colossal permittivity\cite{Lin2014,Benali2022} and interpreted as intrinsic. The present results indicate that this interpretation is incomplete. While comparable hopping barriers persist after dehydration, the dielectric strength collapses after dehydration. This decoupling suggests that Fe$^{+2}$/Fe$^{+3}$ hopping sets the intrinsic relaxation timescale but is insufficient to account for the giant dielectric response.
Further insight can be obtained by comparison with LuFe$_2$O$_4$43, a canonical mixed-valence oxide in which Fe$^{+2}$/Fe$^{+3}$ hopping and charge ordering are intrinsically established. In polycrystalline LuFe$_2$O$_4$ large dielectric dispersion ($\varepsilon' \approx$103–104) is observed, and the characteristic relaxation frequency follows Arrhenius behaviour with an activation energy of $\sim$ 0.29 eV and an attempt frequency of $\sim$1011 Hz. Despite its intrinsic mixed-valence character, the dielectric permittivity of  LuFe$_2$O$_4$ remains bounded and does not reach the colossal values observed in hydrated BiFeO$_3$ ceramics. This comparison highlights a key distinction: Fe$^{+2}$/Fe$^{+3}$ hopping provides the electronic energy scale for thermally activated relaxation but does not by itself produce colossal dielectric permittivity.
Large dielectric enhancements associated with the presence of water have also been reported in several oxide and mineral systems where interfacial polarization develops in heterogeneous structures containing confined moisture\cite{Vasilyeva2014,Haidar1986,Wang2025}. In such cases the dielectric response increases dramatically at low frequency while the characteristic relaxation frequencies remain governed by thermally activated electronic or ionic processes\cite{Samet2022,Samet2015,Jonscher1977}. These observations indicate that the presence of water can strongly amplify the macroscopic dielectric response through interfacial polarization mechanisms without substantially altering the intrinsic activation energies of the underlying charge transport processes\cite{Haidar1986}.
Taken together, these systems illustrate two limiting regimes. Intrinsically correlated mixed-valence oxides establish the electronic activation energy scale yet yield only moderate dielectric permittivity\cite{Ikeda2000,Lafuerza2013}, whereas the presence of confined water in heterogeneous materials can strongly amplify the dielectric response through interfacial polarization effects\cite{Vasilyeva2014,Haidar1986,Wagner1914,Sillars1937}. The present BiFeO$_3$ ceramics lie between these limits. Intrinsic Fe$^{+2}$/Fe$^{+3}$ hopping persists with nearly unchanged activation energies, yet hydration induces a pronounced enhancement of dielectric strength and emergent saddle-point relaxation dynamics.
A closely related phenomenon has been reported in layered clay minerals by Vasilyeva et al.\cite{Vasilyeva2014}, where hydration-controlled saddle-type relaxations appear in the presence of confined water. In those systems the reported water contents are 2.3-5.4 wt\% for kaolinites and 13.7-15.3 wt\% for montmorillonites. In contrast, the present BiFeO$_3$ ceramic holds a total water content below 1 wt\% ($\sim$0.9 wt\%) yet reproduces similar saddle-point relaxation behaviour. BiFeO$_3$ therefore exhibits hydration-controlled dielectric phenomena with nearly fifteen-fold lower water content. This observation suggests that the spatial confinement and interfacial environment of the water molecules, rather than their absolute quantity, may control the magnitude of hydration-induced dielectric amplification.
After dehydration, all relaxation processes revert to simple Arrhenius behaviour with strongly reduced dielectric strengths and the disappearance of the saddle-point curvature. This behaviour is difficult to reconcile with a purely intrinsic electronic mechanism and instead indicates a dielectric response arising from hydration-mediated interfacial polarization superimposed on an intrinsic Fe$^{+2}$/Fe$^{+3}$ hopping background. These observations further suggest that some previously reported colossal dielectric effects in BiFeO$_3$ ceramics\cite{Lin2014,Duong2023} might have in certain cases originated from trace confined hydration.

\section{Conclusion}
We demonstrate that trace, confined water, present at levels below one weight percent, can dominate dielectric relaxation and charge transport in porous BiFeO$_3$ ceramics. Hydration activates a collective dielectric response characterized by an exceptionally large dielectric strength and saddle-point relaxation dynamics, which disappears upon dehydration and are difficult to explain by intrinsic dipolar or defect-controlled mechanisms alone. While intrinsic Fe$^{+2}$/Fe$^{+3}$ hopping sets the underlying relaxation timescale, the colossal dielectric response emerges only when hydration enables correlated interfacial polarization, an extrinsic effect that has long remained unrecognized and has therefore been widely interpreted as intrinsic. A quantitative comparison with layered clay minerals reveals a striking efficiency: BiFeO$_3$ reproduces the same hydration-controlled saddle-point dynamics, with nearly fifteen-fold lower water content ($<$1 wt\% here versus 2-15 wt\% in clays). This establishes confined hydration, rather than water quantity, as the decisive factor governing collective polarization in dense oxides.\\
\textbf{Outlook.} By introducing dehydration-controlled dielectric cycling as a diagnostic approach, this work opens a new avenue for reinterpreting colossal dielectric responses in functional oxides and establishes trace hydration as an active, tunable control parameter rather than an experimental artifact. Controlled re-hydration experiments under defined humidity conditions will provide further insight into the reversibility of the hydration-mediated dielectric response.

	\bibliographystyle{unsrt} 
	\bibliography{references} 

\input{supplementary.tex}
\end{document}

%% file: Supplementary.tex
\section*{Supplementary Material}

\subsection*{S1. Condition for the Saddle-Point Minimum}

The relaxation time is described by
\[
\tau(T) = \tau_0 \exp\left[\frac{H_a}{kT} + C \exp\left(-\frac{H_d}{kT}\right)\right].
\]

Since $\tau > 0$, extrema are determined from $\ln \tau$:
\[
\ln \tau = \ln \tau_0 + \frac{H_a}{kT} + C \exp\left(-\frac{H_d}{kT}\right).
\]

Differentiating with respect to $T$ and setting $d(\ln \tau)/dT = 0$:
\[
-\frac{H_a}{kT^2} + C \frac{H_d}{kT^2} \exp\left(-\frac{H_d}{kT}\right) = 0.
\]

Multiplication by $kT^2$ gives:
\[
C H_d \exp\left(-\frac{H_d}{kT_s}\right) = H_a.
\]

Hence,
\[
\exp\left(-\frac{H_d}{kT_s}\right) = \frac{H_a}{C H_d},
\]
and
\[
T_s = \frac{H_d}{k \ln\left(\frac{C H_d}{H_a}\right)}.
\]

For a real, positive $T_s$, the logarithm must be positive:
\[
\frac{C H_d}{H_a} > 1 \quad \Rightarrow \quad C > \frac{H_a}{H_d}.
\]

Thus, a relaxation-time extremum exists if and only if
\[
C > \frac{H_a}{H_d}.
\]

Now, we examine the second derivative. Let $f = \ln \tau$. Then
\[
\frac{d^2 f}{dT^2} =
\frac{2H_a}{kT^3}
- C \exp\left(-\frac{H_d}{kT}\right)
\left(
\frac{2H_d}{kT^3}
+ \frac{H_d^2}{k^2 T^4}
\right).
\]

Substituting the extremum condition
\[
C \exp\left(-\frac{H_d}{kT_s}\right) = \frac{H_a}{H_d},
\]
we obtain
\[
\left.\frac{d^2 f}{dT^2}\right|_{T_s}
=
\frac{2H_a}{kT_s^3}
- \frac{H_a}{H_d}
\left(
\frac{2H_d}{kT_s^3}
+ \frac{H_d^2}{k^2 T_s^4}
\right).
\]

After cancellation of the first terms,
\[
\left.\frac{d^2 f}{dT^2}\right|_{T_s}
=
\frac{H_a H_d}{k^2 T_s^4} > 0.
\]

Since
\[
\left.\frac{d^2 (\ln \tau)}{dT^2}\right|_{T_s} > 0,
\]
the extremum corresponds to a minimum of $\ln \tau$, and therefore also of $\tau(T)$.

Thus, when the condition
\[
C > \frac{H_a}{H_d}
\]
is satisfied, the saddle-point expression necessarily produces a true relaxation-time minimum at $T_s$.